\documentstyle[aas2pp4]{article}

\def\OVI{\ion{O}{6}}
\def\uhz{$\mu\rm{Hz}$\ }
\def\uhzp{$\mu\rm{Hz}$}
\def\msun{M_{\odot}}
\def\m*{M_*}
\def\lsun{ L_{\odot}}

\def\parcsec{{\tt ''}\mskip -7.6mu.\,}   
\def\simlt{\ {\raise-.5ex\hbox{$\buildrel<\over\sim$}}\ } 

\begin{document}

\title{Asteroseismological Observations of the Central Star of the 
Planetary Nebula NGC 1501}

\author{
Howard~E.~Bond\altaffilmark{1,2},
Steven D.~Kawaler\altaffilmark{3},
Robin Ciardullo\altaffilmark{4},
R.~Stover\altaffilmark{5},
T.~Kuroda\altaffilmark{6},
T.~Ishida\altaffilmark{6},
T.~Ono\altaffilmark{6},
S.~Tamura\altaffilmark{7,8},
H.~Malasan\altaffilmark{7,9},
A.~Yamasaki\altaffilmark{7,10},
O.~Hashimoto\altaffilmark{7,11},
E.~Kambe\altaffilmark{7,10},
M.~Takeuti\altaffilmark{7,8},
T.~Kato\altaffilmark{12,13},
M.~Kato\altaffilmark{12,13},
J.-S.~Chen\altaffilmark{14},
E.~M.~Leibowitz\altaffilmark{15},
M.~M.~Roth\altaffilmark{16,17},
T.~Soffner\altaffilmark{16},
\and
W.~Mitsch\altaffilmark{16}}

\altaffiltext{1}{Space Telescope Science Institute, 3700 San Martin 
Drive, Baltimore, MD 21218; bond@stsci.edu}
\altaffiltext{2}{Guest Observer, Kitt Peak National Observatory, National 
Optical Astronomy Observatories, operated by AURA under contract with NSF}
\altaffiltext{3}{Dept.~of Physics and Astronomy, Iowa State University, 
Ames, IA  50011; sdk@iastate.edu}
\altaffiltext{4}{Dept.\ of Astronomy \& Astrophysics, Pennsylvania State 
University, 525 Davey Lab., University Park, PA  16802; rbc@astro.psu.edu}
\altaffiltext{5}{UCO/Lick Observatory, University of California, Santa Cruz, 
CA  95064; richard@ucolick.org}
\altaffiltext{6}{Nishi-Harima Observatory (NHAO), Sayo-cho, Hyogo 679-53, 
Japan; kuroda@nhao.go.jp}
\altaffiltext{7}{Okayama Astrophysical Observatory, Kamogata-cho, 
Asakuchi-gun, Okayama 719-02, Japan}
\altaffiltext{8}{Astronomical Institute, Tohoku University, Sendai, Japan
980-77; tamura@astroa.astr.tohoku.ac.jp}
\altaffiltext{9}{current address: Bosscha Observatory, Lembang, Bandung
40391, Indonesia; HAKIM@as.itb.ac.id}
\altaffiltext{10}{Department of Geoscience, National Defense Academy, Yokosuka, 
Japan 238; yamasaki@cc.nda.ac.jp}
\altaffiltext{11}{Department of Technology, Seikei University, Musashino-shi, 
Tokyo, Japan 180; hasimoto@aps.seikei.ac.jp}
\altaffiltext{12}{Ouda Observatory, Japan; tkato@kusastro.kyoto-u.ac.jp}
\altaffiltext{13}{Department of Astronomy, Faculty of Science, Kyoto University,
Sakyo-ku, Kyoto 606-01, Japan; tkato@kusastro.kyoto-u.ac.jp}
\altaffiltext{14}{Beijing Astronomical Observatory, Chinese Academy of 
Sciences, Zhongguancun, Beijing 100080, China}
\altaffiltext{15}{Wise Observatory and School of Physics \& Astronomy,
Sackler Faculty of Exact Sciences, Tel Aviv University,
Ramat Aviv, Tel Aviv 69978, Israel; elia@wise1.tau.ac.il}
\altaffiltext{16}{Universit\"ats-Sternwarte M\"unchen, Scheinerstr.~1, 
81679 M\"unchen, Germany; soffner@usm.uni-muenchen.de}
\altaffiltext{17}{current address: Astrophysikalisches Institut Potsdam, An
der Sternwarte 16, 14482 Potsdam, Germany; mmroth@aip.de}

\received{4 June 1996}
\accepted{22 August 1996}

\begin{abstract}

We report on a global CCD time-series photometric campaign to decode the
pulsations of the nucleus of the planetary nebula NGC~1501.  The WC4 central
star is an extremely hot, hydrogen-deficient, ``\OVI''-type object, with some
spectroscopic characteristics similar to those of the pre-white-dwarf
PG~1159$-$035 stars.  NGC~1501 shows pulsational brightness variations of a
few percent with numerous individual periods ranging from 19 to 87~minutes.
The pulsation amplitudes and periods are highly variable, suggesting a
complex pulsation spectrum that requires a long unbroken time series to
resolve.  To that end, we obtained CCD photometry of the central star over a
two-week period in 1991 November, using a network of observatories around the
globe.  We obtained nearly continuous coverage over an interval of almost one
week in the middle of the run.  With this data set, we have identified 10
independent pulsation periods, ranging from 5235~s down to 1154~s. The
pulsation modes changed amplitude significantly during the course of the run,
indicating either real amplitude variations, or that the modes are not fully
resolved over the two-week interval. We find strong evidence that the modes
we see in this star are indeed nonradial $g$-modes.  The ratios of the
frequencies of the largest-amplitude modes agree closely with those expected
for modes that are trapped by a density discontinuity in the outer layers.
This conclusion is strengthened by including single-site observations of this
star, obtained during previous years, in our analysis.  We offer a model for
the pulsation spectrum that includes a common period spacing of 22.30~s and a
stellar rotation period of 1.17~days; the period spacing allows us to assign
a preliminary asteroseismological mass of $0.55\pm0.03 \msun$.  However,
several factors complicate the analysis. Aside from the proximity of the
rotational splitting to 1 cycle per day, this frequency splitting corresponds
closely to {\it period\/} spacings near 20 seconds near the dominant
frequencies of the star.  Thus, the period spacing and frequency spacings are
nearly degenerate.

\end{abstract} 


\section{Introduction}
\setcounter{footnote}{0}

A new and powerful tool for probing the depths of white dwarfs is
asteroseismology.  Stars that are multimode nonradial pulsators can reveal
their internal structure through accurate measurements of the periods of
their photometric (or radial-velocity) variations.  The prototype of the
pulsating hot white dwarfs, PG~1159$-$035 (GW~Vir), has yielded a wealth of
information about its interior, through analysis of its light curve as
obtained by the Whole Earth Telescope (WET); see Nather et~al.\ (1990) for a
detailed description of WET\null.  The detailed analysis of WET observations
of PG~1159$-$035 reported by Winget et~al.\ (1991, hereafter ``WWET'') and
Kawaler \& Bradley (1994, hereafter KB) illustrates what such stars can teach
us; this star shows an unbroken sequence of dozens of pulsation modes that
match expectations from stellar evolution and pulsation theory remarkably
well.  Other GW~Vir stars, though they show fewer modes than PG~1159$-$035,
have also been explored in detail through asteroseismology with WET data
(i.e., PG~2131+066; Kawaler et~al.\ 1995, PG~1707+427; O'Brien
et~al.\ 1997).  In addition, PG~0112+200 shows a rich and decodable pulsation
spectrum (O'Brien et~al.\ 1996).

As illuminating as the seismological study of the four known pulsating
GW~Vir-type white dwarfs has been, planetary-nebula nuclei (PNNs) that
pulsate have an equally enormous potential as probes of stellar evolution.
Nine PNNs are now known to be nonradial pulsators.  The first two to be
discovered were the nuclei of \hbox{K~1-16} (Grauer \& Bond 1984) and Lo~4
(Bond \& Meakes 1990), and six more were discovered in a photometric CCD
survey of \OVI\ and PG~1159 PNNs as summarized by Ciardullo \& Bond (1996).
The total is brought to nine by the discovery of a very faint planetary
nebula around the pulsating PG~1159-type star RX~J2117.1+3412 by Appleton,
Kawaler, \& Eitter (1993).  (In addition, the PNN-like \OVI\ star
Sanduleak~3, although lacking a visible nebula, is a pulsating variable;
Bond, Ciardullo, \& Meakes 1991.)

Unfortunately, PNNs are more difficult to observe than the pulsating PG~1159
stars for several reasons.  One major difficulty is that the pulsation
periods of the PNNs range from about 700~s to more than 5000~s, and thus slow
transparency variations over a given night make disentangling the true
stellar pulsation from variations in the sky transparency difficult.  Another
problem is that, by definition, PNNs are surrounded by nebulosity.  For
photoelectric aperture photometry this reduces the measured pulsation
amplitude, and causes almost intractable problems due to spurious variations
introduced by seeing and small guiding errors.  Still another frustration is
that in the best-studied pulsating PNN to date, \hbox{K 1-16},
a given pulsation frequency
apparently changes significantly from night to night, and changes amplitude
irregularly.  Despite heroic efforts (for example, see Grauer et~al.\ 1987),
the pulsations in \hbox{K 1-16} have still not been fully resolved.  While
such instability could result from intrinsic variations in the star, it could
also result from beating between modes that are closely spaced in frequency,
or from confusion resulting from diurnal aliases or overlapping rotational
splittings (see below).  This seems to be a property of the entire class;
most if not all of the pulsating \OVI\ central stars show complex and
variable behavior (Ciardullo \& Bond 1996).

Most of the observational problems listed above are solved by global
monitoring campaigns with CCD cameras.  CCD imaging systems are capable of
measuring the relative magnitudes of PNNs on time scales as short as
the exposure plus readout time, while the two-dimensional format, which can
include simultaneously--observed comparison stars, avoids the problems
associated with the nebula and variable sky transparency.  In addition, by
using a network of CCD-equipped telescopes around the globe, it is possible to
obtain nearly continuous data on a specific PNN for a week or more, thus
greatly reducing the ambiguities introduced by the daily gaps in single-site
data. 

In this paper, we report the results of a global observing campaign on the
central star of the planetary nebula NGC~1501. NGC~1501\footnote{In this
paper we will refer to central stars of planetary nebulae by the names of the
nebulae.} was first discovered to be a variable star in 1987 by Bond \&
Ciardullo (1993).  It is classified as a WC4 central star, and shows
prominent emission lines of \OVI\ in the stellar spectrum (Smith \& Aller
1969; Heap 1982; Kaler \& Shaw 1984).  Using the HeII Zanstra temperature,
Stanghellini et al. (1994) estimate that NGC~1501 has an effective
temperature of $\log T_{\rm eff}=4.91\pm0.03$; from its visual magnitude and
estimated distance, they obtain $\log L/\lsun=3.33 \pm 0.15$.  Comparison of
its position in the H-R diagram with evolutionary models (see Figure~9 of
Stanghellini et al. 1994) indicates that the mass of NGC~1501 is $\approx
0.56\msun$.  NGC~1501 is very hot, rich in C and O, and hydrogen deficient.
It is similar to the PG~1159 spectral class of pre--white-dwarf stars (see,
for example, Werner et~al.\ 1991; Werner 1995), although evidently somewhat
less evolved.  NGC~1501 pulsates on a time scale of about 25 minutes, with an
amplitude of up to several percent. In general, therefore, it appears to
possess typical properties found in the known pulsating PNNs (Bond
et~al.\ 1993; Ciardullo \& Bond 1996). Because of the spectroscopic
similarities between these stars and the ``naked PG~1159 stars'' it is
presumed that there is a direct evolutionary link, in the sense that the
central stars are the precursors of the PG~1159 stars.  NGC~1501 is thus an
early link in the chain of evolution between stars with active nuclear
burning (AGB and post-AGB stars) and stars that are destined simply to cool
and fade (white-dwarf stars).

NGC~1501 is probably the best northern PNN candidate for a global monitoring
campaign, since it is relatively bright (broadband optical magnitude of about
14), and because its northerly declination ($+61\arcdeg$) makes it observable
for nearly 11~hours per night from a single northern-hemisphere observatory
during November.

We are entering into a very exciting time in asteroseismology and in the
study of PNNs.  The NGC~1501 results presented in this paper represent the
first detailed analysis of the nonradial pulsations of a PNN\null. 
In the next section, we describe the organization of our CCD network for the
NGC~1501 campaign, the observing procedures, and the resulting time-series
data. We then describe the results of our analysis of the 1991 campaign data,
and the pulsation spectrum of NGC~1501.  With the aid of our older archival
data on NGC~1501, we then construct an initial model of the underlying
pulsation spectrum, and obtain the stellar mass and rotation period.  We
conclude with an attempt to understand this pulsation spectrum in terms of
models of PNNs and PG~1159 stars.

\section{Observations}

\subsection{CCD Photometry}

For 14 days in 1991 November, we monitored the central star of NGC~1501 with
0.8- to 1.0-m telescopes at five sites around the globe: Kitt Peak National
Observatory (KPNO) and Lick Observatory in the United States, Okayama
Astrophysical Observatory in Japan, Wise Observatory in Israel, and
Mt.~Wendelstein Observatory in Germany.  (Four additional sites also
attempted to observe the object during the campaign, but due to weather and
technical problems, their contributions could not be used.) At each site, the
object was observed throughout the night via a series of short ($\sim 2$
minute) exposures with a CCD detector, and a filter which accepted visual or
red light.  In all cases, the field of view of the CCD was large enough to
include NGC~1501 itself and three nearby comparison stars (shown in
Figure~1).  A summary of the telescopes and observing setups used in this
campaign is presented in Table~1.  Table~2 details the active observing time
of each observatory.  Good weather prevailed at many of the sites, so that
between 1991 November 18 and November 28, useful data were obtained for 173
hours out of a possible 216.  For example, all nine scheduled nights at KPNO
produced usable data (albeit through varying amounts of thin cirrus and haze,
which are of little concern for differential CCD photometry).  It may be
noted that the campaign was centered around full moon, which is not a problem
for CCD observations of a star this bright, and made it easier to obtain
telescope time.

Our data-reduction techniques are described in detail by Ciardullo \& Bond
(1996), but are summarized here.  Since each night of observing produced
between 200 and 300 CCD images, a frame-compaction algorithm was first used
to reduce the data to a manageable size and format.  After debiasing and
flat-fielding the data, we extracted $32 \times 32$-pixel regions surrounding
NGC~1501 and its three comparison stars from each individual frame.  [Because
of the small ($0\parcsec 2$/pixel) plate scale of the Okayama observatory
data, the frames from this telescope were first binned $4 \times 4$ before
the object extraction took place.] The extracted regions were then moved into
a set of ``packed'' pictures, each containing data from 16 original frames.
Although this operation eased the burden of data handling, it in no way
affected the subsequent data reductions, since the data from each original
frame were always analyzed independently.

Photometric reductions were then accomplished using DAOPHOT algorithms
(Stetson 1987) inside IRAF\null.  Each frame's point-spread function (PSF)
was first defined using the three pre-selected comparison stars.  The
relative magnitudes of these stars and that of NGC~1501 were then computed
using the DAOPHOT PSF-fitting algorithm with a fitting radius of the order of
the image full-width at half-maximum (FWHM)\null.  (Because NGC~1501's nebula
is of relatively low surface brightness, no special image modeling of the
nebula was needed [cf.~Ciardullo \& Bond 1996]).  From these data, NGC~1501's
relative intensity was measured by differencing the raw magnitude of NGC~1501
and the flux-weighted mean instrumental magnitudes of the comparison stars.

Because of its favorable declination ($+61^\circ$), each telescope in our
network could observe NGC~1501 for over 10 hours a night.  However, during
these long runs, the airmass of the object changed substantially.  This
change, coupled with the color difference between the PNN and our comparison
stars (see Table~3 for coordinates and our measured magnitudes and colors),
could, in principle, introduce an apparent long-term variation in NGC~1501's
brightness, especially in data taken through a broadband filter.  We
therefore checked for this effect by regressing NGC~1501's computed
instrumental magnitude against airmass for each observatory's data, and
measuring the slope of the relation.  In all cases, the systematic error was
$\simlt 0.015$~mag per airmass, with a one-sigma uncertainty of about half
this value.  Since the effect is small, and in any case the time scales of
primary interest are significantly shorter than the individual runs, we did
not make any correction for differential extinction between NGC~1501 and the
comparison stars.

A second correction to NGC~1501's instrumental magnitude comes from the effects
of seeing.  By restricting our photometry to apertures of the order of the
FWHM and treating each frame identically, any error caused by mis-estimating
the sky plus nebula surrounding NGC~1501 should have been minimized.
However, under conditions of variable seeing, small systematic errors in the
relative magnitudes can still exist.  To investigate this possibility, we
again performed a least squares fit, this time regressing NGC~1501's
differential magnitude against the FWHM of the each frame's PSF\null.  Once
again, the seeing correction proved to be small, typically less than
$0.01$~mag per arcsec.  As a result, we did not correct for this in the final
analysis.

\subsection{Time-Series Analysis}

After obtaining the intensity of NGC~1501 on each individual frame, we
computed the ``modulation intensity'' by subtracting off, and then
dividing by, the star's overall mean intensity.  After this was done,
individual points that were far away from the rest (generally due to
cosmic-ray hits on the variable or one of the comparison stars) were edited
out, and the observation times were reduced to Barycentric Julian Dynamical
Date (BJDD)\null.  The data from each site were reduced separately, and then
combined into a final light curve.  We made no effort to weight the data for
the different apertures of the telescopes, CCD sensitivities, or other 
site-dependent effects.

A subset of the data is shown in Figure~2 in 24-hour strips for seven days at
the heart of the run.  Figure~2 includes only data from Okayama, KPNO, Wise,
and Wendelstein observatories.  (Because the Lick data largely overlap those
from KPNO, including these data would make the plot more complicated.
However, the agreement between observatories is quite good.  In the analysis
below we do include the Lick data when determining accurate frequencies.) The
principal variation of about 1200 to 1500 ~s is readily visible in this
plot.  The plot also shows the temporal coverage of our data; the lack of any
long vertical stripes without data is an indication that there were no
strictly periodic gaps in this segment of the light curve.  The window
function for this run, obtained by sampling a single-frequency sine wave of
fixed amplitude at all of the observed sample times, is shown in Figure~3.
The window function has 1 cycle day$^{-1}$ alias peaks at several percent 
of the power of the main peak.

We computed the Fourier transform of the reduced data set using a brute-force
discrete Fourier transform, and squaring the amplitude of the complex
transform to produce the power spectrum.  This gives the total modulation
power (mp) as a function of frequency.  The square root of the modulation
power gives the modulation amplitude (ma).  In many cases, we find that plots
of modulation amplitude (which we call ``amplitude spectra'') are more useful
than power spectra, as the amplitude spectra de-emphasize strong peaks.  The
representation of pulsation power and amplitude in units of mp and ma is
discussed in Winget et~al.\ (1994).  The amplitude spectrum is shown in
Figure~4 for the entire frequency range of interest; for comparison, the
window function is plotted again, below the strong peak at 866~$\mu$Hz.

While the amplitude spectrum shows relatively clean peaks, in some cases the
(small) alias sidelobes affect the perceived amplitudes and frequencies of
lower-amplitude modes.  Essentially, the low-amplitude peaks can be shifted
away from their true frequencies by power that leaks from the
larger-amplitude peaks.  Also, the amplitude spectrum does not give a
quantitative estimate of the uncertainty in the frequency, amplitude, or
phase of a given periodicity.  We therefore used a linear least-squares
fitting of the time series photometry to determine the precise frequencies,
amplitudes, and phases of many peaks simultaneously.  Careful use of least
squares allows these quantities to be determined accurately for all modes
included in the fit (i.e., Kawaler et~al.\ 1995; O'Brien et~al.\ 1996); the
least-squares values are unaffected by the sampling pattern if all modes are
included in the fit.  This procedure also provides a formal estimate of the
uncertainty of these quantities.

We used the amplitude spectrum of the entire data set to provide initial
guesses for the 11 clearest peaks for use in the initial least-squares fit.
The least-squares frequencies match the peaks in the amplitude spectrum
within the errors.  These frequencies are listed in Table~4.  The times of
maxima are with respect to $T_0 = \rm BJDD\,\,2448571.50367$, which
corresponds to a barycentric date of 1991 November 14.0.  The only suspicious
peak identified in the power spectrum is the peak at 855.5~\uhzp, which
appears to be close to the $-1$ cycle d$^{-1}$ alias of the large peak at
866.3~\uhzp.  Because of its relatively large power, and the difference in
frequency between it and the true $-1$ cycle d$^{-1}$ alias, we suspect that
it might be a real periodicity in the star.  Because of our uncertainty,
however, we exclude it in the following analysis.

\subsection{Archival Data for NGC 1501}

Since only 10 modes were definitively identified in the 1991 global campaign,
these data alone are insufficient for the kind of asteroseismological
analysis performed on the GW~Vir stars.  However, other pulsating white
dwarfs show modes that appear and disappear on relatively long time scales;
when new modes appear, they appear at frequencies expected given the fact
that they are $g$-modes.  That is, the new modes usually fit a pattern of
equal period spacing and/or rotational frequency splitting.  In the case of
PNNs, Grauer et~al.\ (1987) and Ciardullo \& Bond (1996) show that the modes
seen in PNNs can come and go on much shorter time scales than in GW~Vir
stars.  If these modes are indeed normal modes of oscillation, they must
occur at the limited set of frequencies fixed by the structure of the star.
Therefore, analysis of data taken at earlier seasons could reveal additional
frequencies which can be used in the analysis.

In the hopes of finding such additional modes, we reexamined time-series
photometry of NGC~1501 obtained by Ciardullo \& Bond (1996) at KPNO between
1987 and 1990. The individual observing runs are listed in Table~5.  Data
were obtained using the same CCD system as in the present study, and were
reduced following the same procedures as for the 1991 global campaign.
Because these were all single-site runs, the power spectra of these
additional data are riddled with diurnal aliases.  However, in several of the
seasons, a few large-amplitude modes can be identified unambiguously. The
frequencies, periods, and amplitudes of the dominant modes in those runs are
listed in Table~6.

We note that these data sets show abundant evidence for additional
periodicities, but the aliasing problem is severe.  To disentangle aliases
from the true frequencies we followed the prewhitening and synthesis
procedures of O'Brien et~al.\ (1996).  In some cases we were able to make
confident frequency determinations for these runs.  In Table~6, the most
insecure frequency determinations are indicated with colons (:).  A few
points are readily apparent.  First, the power present in the dominant modes
from these runs is much larger than in the 1991 campaign.  Moreover, the
period with largest amplitude changes from year to year; only in the 1991
data is the 866~\uhz mode among the largest.  Finally, it is clear that
NGC~1501 displayed a number of modes in the recent past that were not present
during the 1991 campaign.  These additional modes can be used to constrain
models of the underlying pulsation spectrum of the star.

\section{Frequency and Period Patterns in Nonradial $g$-modes}

NGC~1501's pulsation frequencies, if caused by normal nonradial $g$-modes,
should show some systematics.  It is these systematics which allow us to
place seismological constraints on the global parameters of the star.  This
section briefly describes some of the systematics of nonradial $g$-modes that
are present in pulsating white dwarfs and that we might expect in the power
spectrum of NGC~1501.

\subsection{Frequency Splitting}

One of the observed properties of nonradial oscillations is that equal
frequency spacings are often seen when ``bands'' of power are resolved.  For a
perfectly spherically symmetric star, the frequencies of all nonradial modes
with a given degree $\ell$ and order $n$ are identical.  However, departures
from spherical symmetry (which can be caused by rotation, magnetic fields, or
some combination thereof) can lift the degeneracy.  The most common example is
rotation, which splits the degenerate modes into $2\ell+1$ components (if all
values of the azimuthal index $m$ are present) as follows:
\begin{equation}
\nu_{nlm}=\nu_{nl0}-m \frac{1}{P_{\rm rot}}(1-C_{nl}),
\end{equation}
where for white dwarf stars, to within a few percent,
\begin{equation}
C_{nl}=\frac{1}{\ell(\ell+1)}
\end{equation}
(Brickhill 1975; WWET 1991).  Thus, if NGC~1501 rotates, we can expect to see
triplets split by one half of the rotation frequency for $\ell=1$ modes.  We
follow the convention that the lower-frequency component of a triplet is
identified as the $m=+1$ component, and the higher-frequency component is
labeled as the $m=-1$ component.  Such equally spaced (or nearly equally
spaced) triplets are commonly seen in pulsating PG~1159 stars and in the
cooler DB and DA white-dwarf pulsators.

\subsection{Period Spacings}

In the limit of high overtones (that is, when the number of radial nodes is
much greater than $\ell$) $g$-modes are approximately equally spaced in
period for a given value of $\ell$ and consecutive values of $n$, i.e., 
\begin{equation}
\Pi_{nl}=\frac{\Pi_0}{\sqrt{\ell(\ell+1)}}(n + \epsilon),
\end{equation}
where $\epsilon$ is a small number.  The period spacing ($\Pi_0$) is
determined largely by the mass of the star, and is only weakly dependent on
the stellar luminosity and surface compositional stratification.  KB, 
and references therein, discuss how the value of this period
spacing is used to constrain the mass of GW~Vir stars, and how departures
from uniform spacing allow the determination of subsurface structure.  Since
the pattern of equally spaced (in period) multiplets has been seen in several
of the GW~Vir stars with period spacings of approximately 20-23 seconds, we
might expect a similar spacing in the periods of NGC~1501.

The relatively small number of observed modes in NGC~1501, coupled with the
long periods and resulting lack of {\it period\/} resolution at the frequency
resolution set by the run length, makes finding period spacings a difficult
task using the 1991 data alone.  As we show in the next section, the archival
data allow us to identify a possible period spacing.

\subsection{Strange Frequency Ratios}

WWET and KB found that the period spacings seen in
PG~1159 itself were not precisely equally spaced.  They noted that this could
result from mode trapping; that is, a subsurface composition transition
region can cause a density discontinuity which acts as a reflecting boundary
between the surface and the deep stellar interior.  KB
identified this layer as the transition zone between the helium-rich surface
layer and the carbon/oxygen core.  In their analysis, KB 
computed so-called trapping coefficients, which are related to the frequency
ratio of trapped modes.  If trapped modes are identified in a star, then the
period ratio between the modes can be compared with these coefficient
ratios.  In fact, the dominant peaks in the amplitude spectrum of the
pulsating central star RX~J2117.1+3412 show period ratios nearly exactly
equal to $\sqrt{3}/2$ (Vauclair et~al.\ 1996), which is compatible with the
coefficients computed by KB.  The precision with which
the period ratios in RX~J2117.1+3412 match $\sqrt{3}/2$ is remarkable, much
better than~1\%.  This suggests that dominant peaks in RX~J2117.1+3412 are
indeed trapped modes.

It is immediately apparent that the same ratio exists in NGC~1501.  The
frequency ratio between the peaks at 866~\uhz and 758~\uhz is 0.8755, or 1.1\%
larger than $\sqrt{3}/2$.  The next two peaks in the sequence would be at about
656~\uhz and 569~\uhzp.  We do see a small (but still significant) peak at
661~\uhz (at a ratio 0.7\% smaller than $\sqrt{3}/2$) and a large peak
at 568~\uhz (at a ratio 0.8\% bigger than $\sqrt{3}/2$).  These observed ratios
are all much closer to to $\sqrt{3}/2$ than seen in the theoretical models
of KB.  Apparently the pulsating central stars NGC~1501
and RX~J2117.1+3412 are better calculators of $\sqrt{3}/2$ than are the computer
models!

It is very interesting to see that a large-amplitude mode exists in past data
at 656-657~\uhzp. Figure~5 shows the spectrum of the 1988 data, along with a
representative window function. This frequency falls exactly in the
$\sqrt{3}/2$ sequence ratio; the next frequency expected in the sequence
below 758~\uhz is 656.8~\uhzp, and the next frequency expected above
567.9~\uhz is 655.8 \uhzp.  Thus the archival data strongly support the
identification of a series of periodicities with frequencies near the
``magic'' ratio of $\sqrt{3}/2$.

\section{Decoding the Pulsation Spectrum of NGC~1501}

\subsection{Identification of Rotational Splitting}

The 656~\uhz periodicity that dominated much of the archival data is not
present in the 1991 data, but there is a modest peak at 661.1~\uhzp.  The
frequency difference here is about 5.4~\uhzp.  In the 1987 data and the
second half of the 1989 data sets, there are clear peaks at 668.0~\uhz and
666.3~\uhzp, respectively.  With the 661~\uhz peak from the 1991 campaign,
these three peaks, seen in different years, may all be components of a
uniformly split triplet with $\delta\nu=10.2$~\uhzp.  Triplet structures like
this are seen in many of the pulsating GW~Vir stars; as described in the
previous section, it is the expected signature of rotational splitting of
$\ell=1$ modes.  Armed with evidence of a frequency triplet centered near
661~\uhzp, we used the archival data and campaign data to reveal other
evidence for frequency splittings of approximately 10~\uhzp.

While the 1991 data show an unaliased peak at 567.9~\uhzp, the strongest peak
in the runs kp89-01 and kp89-02 is clearly at 558.0~\uhzp.  This same
frequency is present in 1988 data.  Figure~6 shows this region of the power
spectrum of the 1988 runs.  To try to decompose this portion of the spectrum
in the presence of 1 cycle/day aliases, we follow the technique of
power-spectrum synthesis described in O'Brien et~al.~(1996).  This procedure
allows us to find the minimum number of independent frequencies that best
reproduce the observed power spectrum.  In the top two panels of Figure~6, we
show the observed power spectrum as a thick line.  In these panels, a thin
line shows a synthetic power spectrum created using two sinusoids with
amplitudes, phases, and frequencies obtained by least-squares fits to the
light curves.  The two frequencies used in each synthesis are marked in the
figure. The rest of the peaks in the panel arise solely from the window
patterns associated with the input frequencies and gaps in the data; no noise
was added in this simulation.  In the topmost panel of Figure~6, the
higher-frequency peak has a frequency close to that seen in the 1991 data
($\approx 568$~\uhzp).  The fitting and synthesis procedure reveals that the
two frequencies do not adequately reproduce the observed power spectrum,
indicating that either one or both frequencies are incorrect, or that more
frequencies are present.  However, if we fit the light curve by setting the
higher frequency at $\approx 558$~\uhz (second panel), the entire region is
very well matched.  Based on this agreement, we are confident that our
identification of a 558~\uhz periodicity in the 1988 data is correct.  Thus
there is a pair of peaks separated by about 10~\uhz centered at 562~\uhzp.
This is (to within the frequency errors) entirely consistent with the
frequency spacings in the 656-666~\uhz region.

The archival data support two additional frequency bands with the identified
splitting near 530~\uhz and 700~\uhzp.  The 1991 data show a peak at
528.3~\uhzp; the archival data show a peak at 533.5~\uhz (in 1988;
cf.~Figure~6), and at 532.5~\uhz (in 1989).  Thus, we have evidence for a
pulsation mode approximately 5~\uhz higher than in 1991.  Though there is no
evidence of a third peak in this frequency range, the data are consistent
with a frequency splitting of modes with the same $\ell$ and $n$ but
consecutive values of $m$.  In 1988, 1989, and 1990 a strong peak was also
evident at a frequency of 708~\uhzp; this peak lies almost exactly 10~\uhz
above the 698.6~\uhz peak seen in 1991.

The mean value of the rotational splitting between the $m=+1$ and $m=-1$
modes in NGC~1501 is 9.9~\uhz; equations (1) and (2) then imply that NGC~1501
rotates with a period of 1.17~days.  If this 9.9~\uhz splitting is truly
present in NGC~1501, then it would explain why this is such a difficult star
to analyze from a single site!  The 1 cycle/day alias separation is
11.6~\uhzp; data from a single site would need to cover at least 11 days to
begin to resolve the true power from the alias.  Since the modes present in
NGC~1501 appear to come and go on this same time scale, identification of the
true peaks from a single site is nearly impossible.

\subsection{Period Spacings}

While the number of modes is still a bit small, we can use the measured
frequencies in the archival runs, along with the 1991 global campaign, to try
to model all of the observed frequencies in terms of modes that are equally
spaced in period and rotationally split in frequency with a total splitting
of $\approx 10$~\uhzp.  Unfortunately, to search for period spacings, we need
to identify the $m=0$ components of the triplets in those cases where all
three modes are not present.  Complicating matters further, the period
spacings are not expected to be precisely uniform, as mode trapping by a
subsurface composition transition region could cause small variations in the
period spacing with period, as in the pulsating PG~1159 stars (KB).

To compute the period spacings, we begin by determining the periods of the
$m=0$ modes.  The triplet near 656~\uhz tells us that the $m=0$ central
frequency can be estimated when we see two modes separated by 10~\uhzp.  From
the frequencies in Tables~4 and 6, we see that $m=0$ periods are available
for the 563~\uhz group, the 661~\uhz group, and the 703~\uhz group.  The
differences in these periods are 263~s and 91~s; these should share a common
multiple which itself is a multiple of the intrinsic period spacing.  The
ratio of these is 2.89; that is close enough that we will call it 3; thus the
period spacing is either approximately 90 seconds or some integer factor
(i.e., 45, 30, 22.5, 18, or 15 seconds).

To reduce this degeneracy, we can appeal to the theoretical calculations of
periods and period spacings in PG~1159 stars by KB.
Assuming that these are $\ell=1$ modes, and that NGC~1501 is a normal
planetary-nebula central star, we can eliminate large values for the period
spacing. Values above 25~s for the period spacing would imply a mass $M <
0.5\msun$, well below the minimum mass needed for a single C/O white-dwarf
star.

We next consider the mode at 758.5~\uhz seen in the 1991 data.  With no other
nearby frequency seen in any of the archival data we cannot claim an
identification of $m$.  If it is $m=0$, then the period is 104 seconds
shorter than for the $m=0$ component of the 703~\uhz group.  This is not
commensurate with the other two period intervals.  If this mode is actually
$m=-1$, then the $m=0$ component would have a period 94.9~s longer than the
703~\uhz group---which is also not commensurate.  However, if the $m=0$
component is 5~\uhz higher in frequency (i.e., at a frequency of 763.5~\uhzp), 
then the $m=0$ period for this mode is 110~s shorter than for the 703~\uhz 
group.  This number is compatible with a period spacing of 22~s.

Since an integer number of spacings must separate the periods of the modes,
the difference between 1776 seconds and 1312 seconds corresponds to 21
intervals of 22.1~s.  The subintervals between successive $m=0$ modes are 12
intervals of 21.9~s, four intervals of 22.8~s, and five intervals of 22.0~s.
With a refined estimate of the period spacing interval, we can continue this
analysis and attempt to identify values of $m$ and the number of intervals
between successive modes ($\Delta n$) for each of the frequencies seen in the
1991 data set (and several from the archival runs as well).  Finally, a
least-squares fit of the periods and $\Delta n$'s allows identification of the
period spacing.  This spacing is 22.30 seconds.  Table~7 presents this model
for the pulsation spectrum of NGC~1501. In all cases, the model matches the
observed periods to within 4~s.  Variations this large (and larger) are
present in the pulsating PG~1159 stars that have been studied by WET, so the
model appears to be satisfactory.

Using equation (5) of KB, and noting that this
equation contains a sign error on the exponent of the luminosity term, we can
use the 22.30~s period spacing to estimate the mass of NGC~1501 as $0.55 \pm
0.03 \msun$. The quoted uncertainty reflects the fact that the luminosity of
NGC~1501 is uncertain; while Stanghellini et al. (1994) estimate an
uncertainty of only about 35\%, we allow that the luminosity could be
anywhere between $\approx 5000\lsun$ and $\approx 100\lsun$ to retain a
conservative error estimate in our mass determination. In addition, KB
derived the relationship between mass and period spacing for
PG~1159 stars, which are more highly evolved than NGC~1501; the period spacing
does tend towards an asymptotic value at high luminosities in their Figure~2,
however.

\section{Summary and Conclusions}

With a coordinated global observing campaign, we have obtained a
high-resolution power spectrum of the pulsations of the central star of the
planetary nebula NGC~1501.  The star showed 10 clear periodicities during
this observing run, with periods ranging from 5200~s down to 1154~s.  The
largest-amplitude pulsations occur between 1154~s and 2000~s.  Additional
data from prior years reveal that NGC~1501 changes its pulsation properties
dramatically over month to year time scales.  Past observations have found it
pulsating with amplitudes twice as large as seen in 1991.  The pulsations
present in prior years in some cases occur at the same frequencies as in
1991, but in other cases the pulsations appear at different (but not random)
frequencies.  We note that as a PNN, NGC~1501 must be evolving relatively 
rapidly; with many modes seen at the same frequencies in subsequent 
years, we are optimistic that further analysis of the data may yield a 
measurement of the evolutionary change in one or more of the stable modes.

Most of the pulsation frequencies are accounted for by a simple model which
includes just two input parameters. We attribute the 9.9~\uhz spacing
exhibited by several multiplets to rotational splitting, implying that the
star's rotation period is 1.17~days. Moreover, there is a nearly uniform
underlying period spacing of 22.30~s between the various multiplets, implying
an asteroseismological mass for NGC~1501 of $0.55\pm0.03\msun$.  This mass is
in close agreement with the mass determined from its position in the H--R
diagram and comparison with evolutionary models (i.e., Stanghellini et al.
1994).  Despite its unusual spectrum, then, NGC~1501 is a garden-variety PNN
in terms of its mass; the mass we find is similar to other estimates of PNN
masses.  It also is consistent with the star being a progenitor of the
PG~1159 spectroscopic and pulsation class.

Our success in analyzing NGC~1501 is muted by the significant observational
difficulties presented by this star.  NGC~1501 changes its pulsation behavior
on time scales that prevent detailed monitoring.  A global campaign is
required to ensure that the 10~\uhz rotational splitting is resolved from
diurnal aliases, but maintaining such a vigil over the months necessary to
look for changes in pulsation frequencies and amplitudes would be an enormous
task.  Moreover, in addition to the alias problem, there is an unfortunate
degeneracy in the pulsation spectrum of this star.  A period spacing of 22.30
seconds corresponds to a frequency difference of 10~\uhzp, when the periods
are near 1500~s (i.e.,  at frequencies near 666~\uhzp).  Thus the
rotationally split low-frequency components of the power spectrum overlap
with the consecutive overtone spacing for periods of 1500~s and longer.  This
may in part account for the difficulty faced in resolving this star, but
unlike diurnal aliasing (which also introduces similar frequency splitting),
observing from around the world does not help.

NGC~1501 is but one of nine PNNs that are now known to show nonradial $g$-mode
pulsations (Ciardullo \& Bond 1996).  Data have already been obtained for
RX~J2117.1+3412  by the WET collaboration, and are currently being analyzed. 
We also have data on Sanduleak~3 that are currently being examined.  With this
paper's demonstration that NGC~1501 is understandable in terms of well-known
principles of asteroseismology, it is our hope that the similarities between
it and the other members of the class will accelerate analysis and
interpretation of their pulsations.  The seismological study of this fleeting
stage of stellar evolution has now begun. 

\acknowledgments

HEB thanks Kitt Peak National Observatory for outstanding equipment and
support, and thanks the participants for their patience during the analysis
of our data. Drs.~R.~Kudritzki and M.~Takeuti were instrumental in setting up
collaborations with German and Japanese observers, respectively.  SDK
acknowledges the National Science Foundation for support under the NSF Young
Investigator Program (Grant AST-9257049) and Grant AST-9115213 to Iowa State
University. RC also acknowledges an NSF Young Investigator Grant.  EML
acknowledges the Israeli Academy of Sciences for supporting astronomical
observations at the Wise Observatory.

\newpage

\figcaption[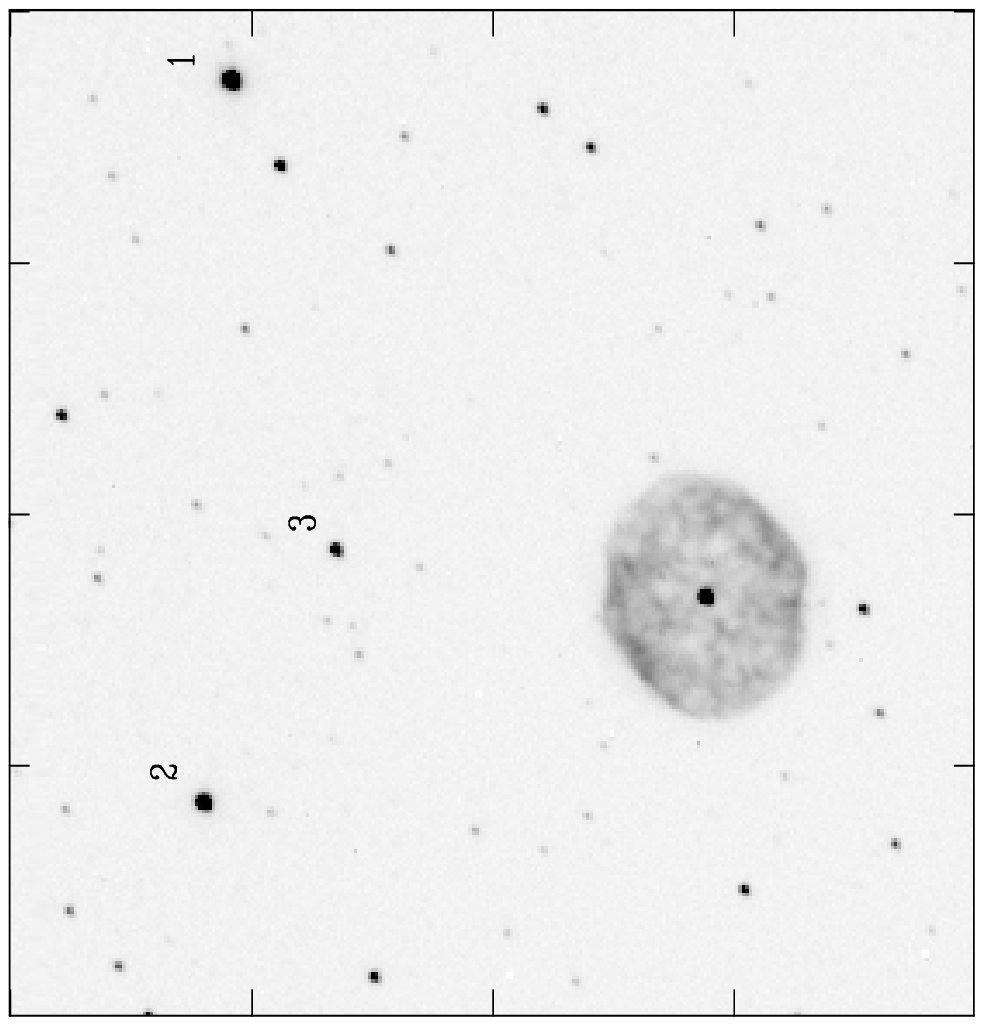]{Sample full--frame image of NGC 1501 and three
comparison stars, obtained through a Str\"omgren $y$ filter with the KPNO
0.9-m telescope. North is at the top, east is on the left, and the image is
approximately $4\farcm3$ wide.}

\figcaption[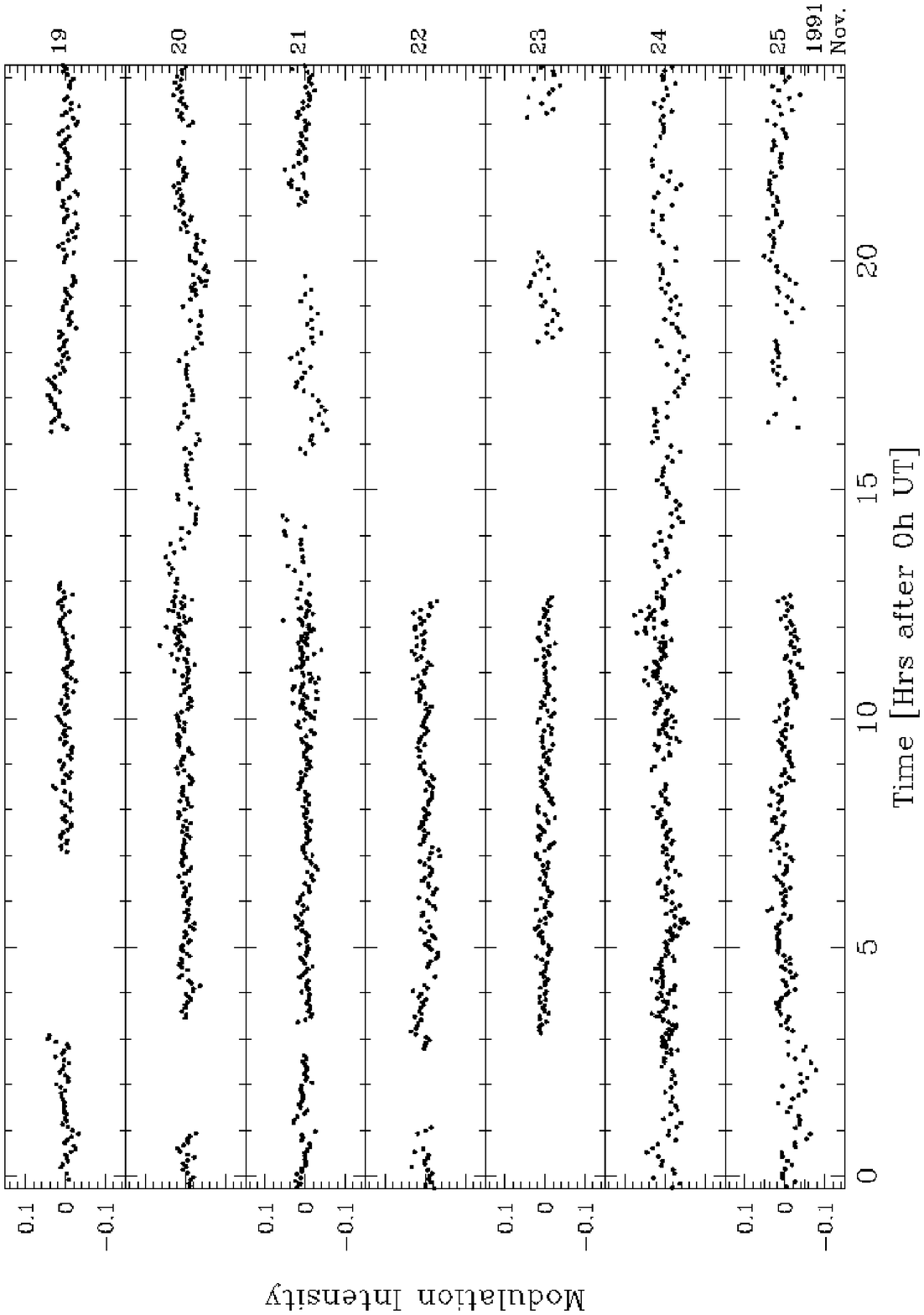]{Combined light curve of NGC~1501 for seven days
during the middle of the 1991 global campaign. Each horizontal panel
represents 24 hours of data, with the UT date indicated to the right of the
panel.  Data from KPNO, Okayama, Wise, and Wendelstein are included in this
plot; since the data from Lick overlap the KPNO data almost exactly, those
data points were not plotted to retain clarity. Data from about 3 to 13 hours
UT are from KPNO; from 10 to 20 hours are from Okayama; from 17 to 2 hours
are from Wise; and from 20 to 5 hours are from Wendelstein.}

\figcaption[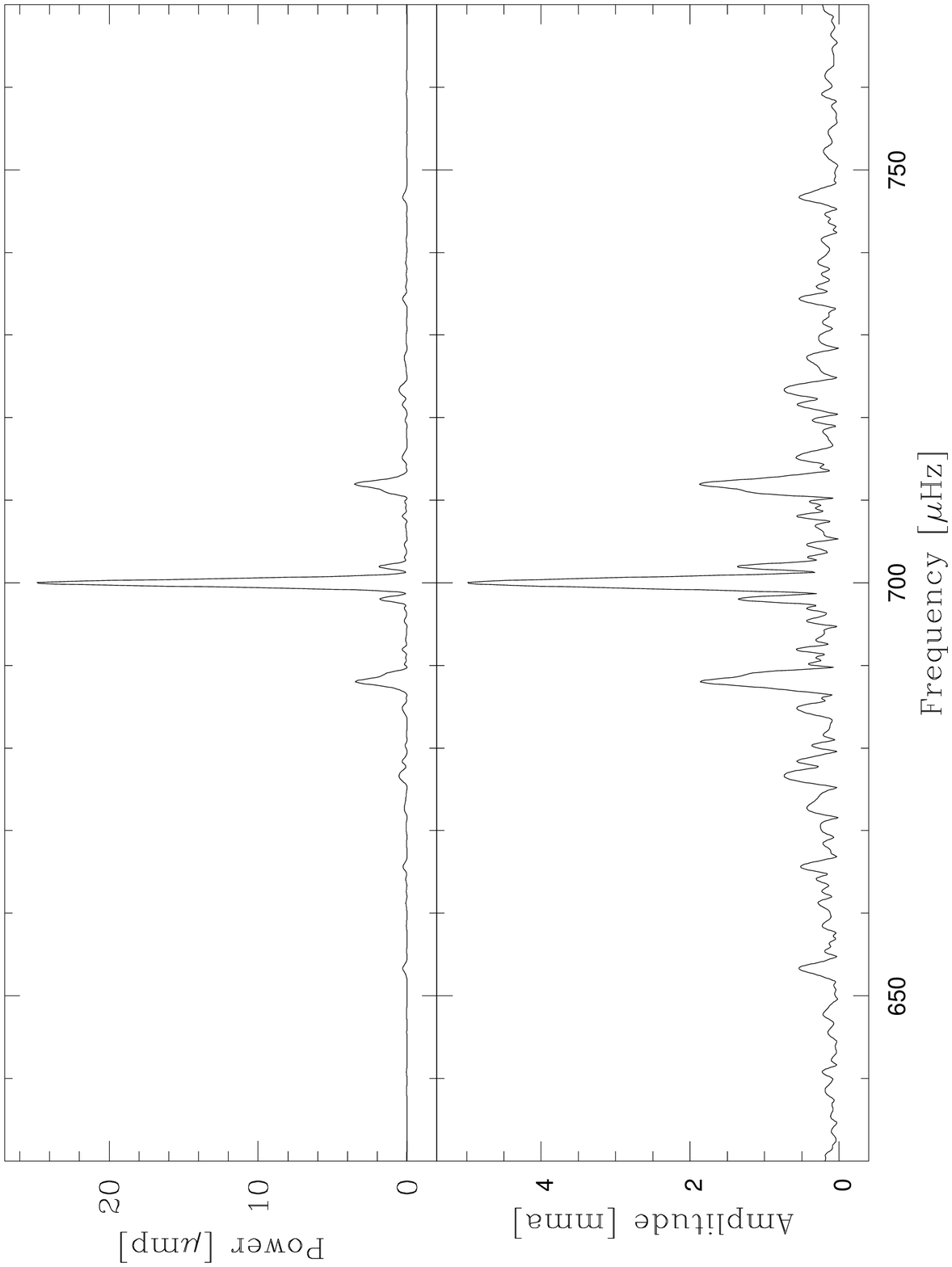]{The spectral window of the global campaign,
computed by sampling a noise--free single--frequency sinusoid at the same
times as present in the data.  The top panel shows the power spectrum of the
window function, while the bottom panel shows the amplitude spectrum.}

\figcaption[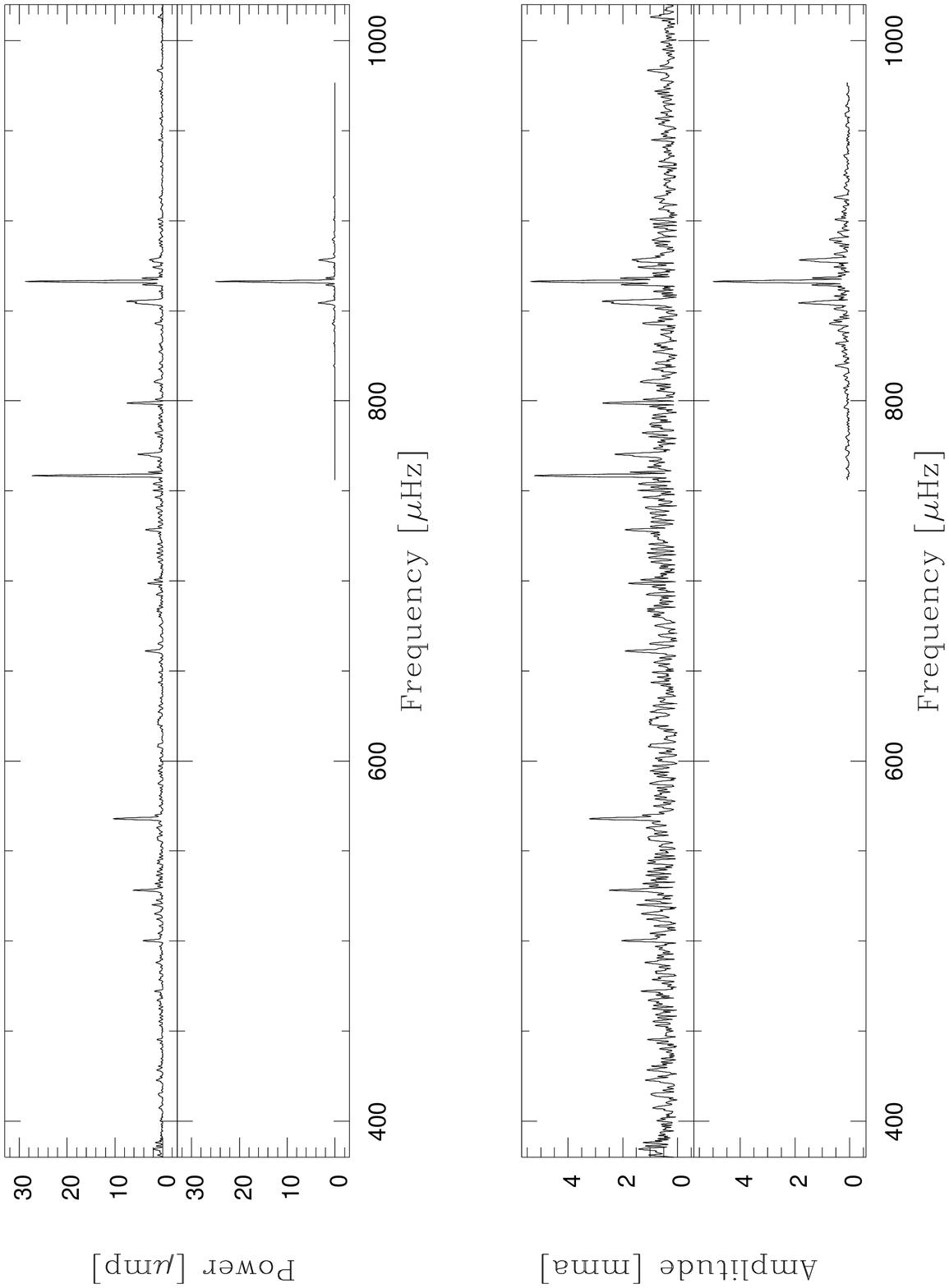]{The power spectrum and amplitude spectrum of the
combined data from the global campaign.  The top panels show the power
spectrum (and the window function) while the bottom panels show the amplitude
spectrum (and the corresponding window function).}

\figcaption[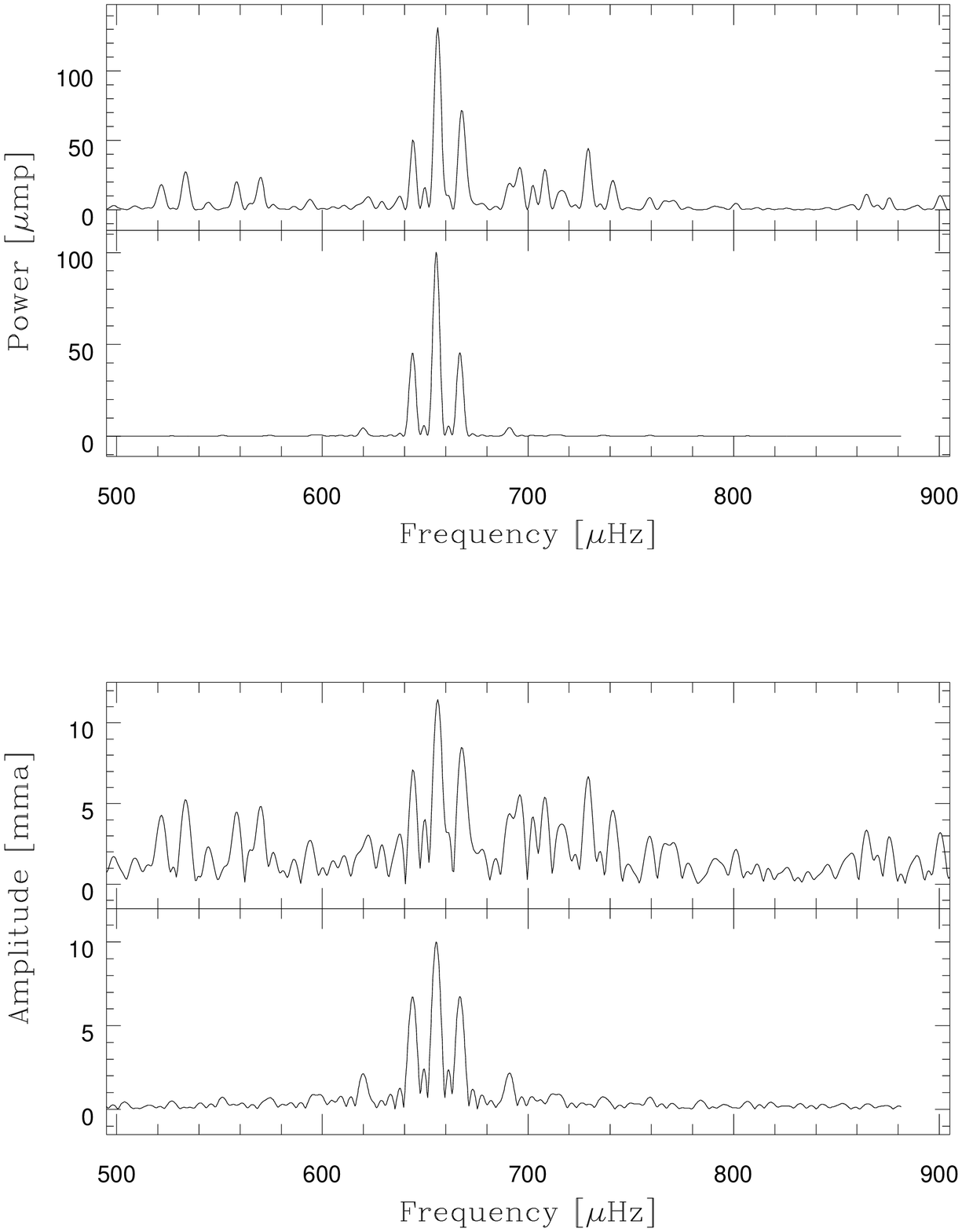]{The power and amplitude spectra for the central
star of NGC 1501, as observed at Kitt Peak on three consecutive nights in
1988 November.  The spectra and windows are as in Figure 4.  Note that the
dominant peak lies at 656~\uhzp.}

\figcaption[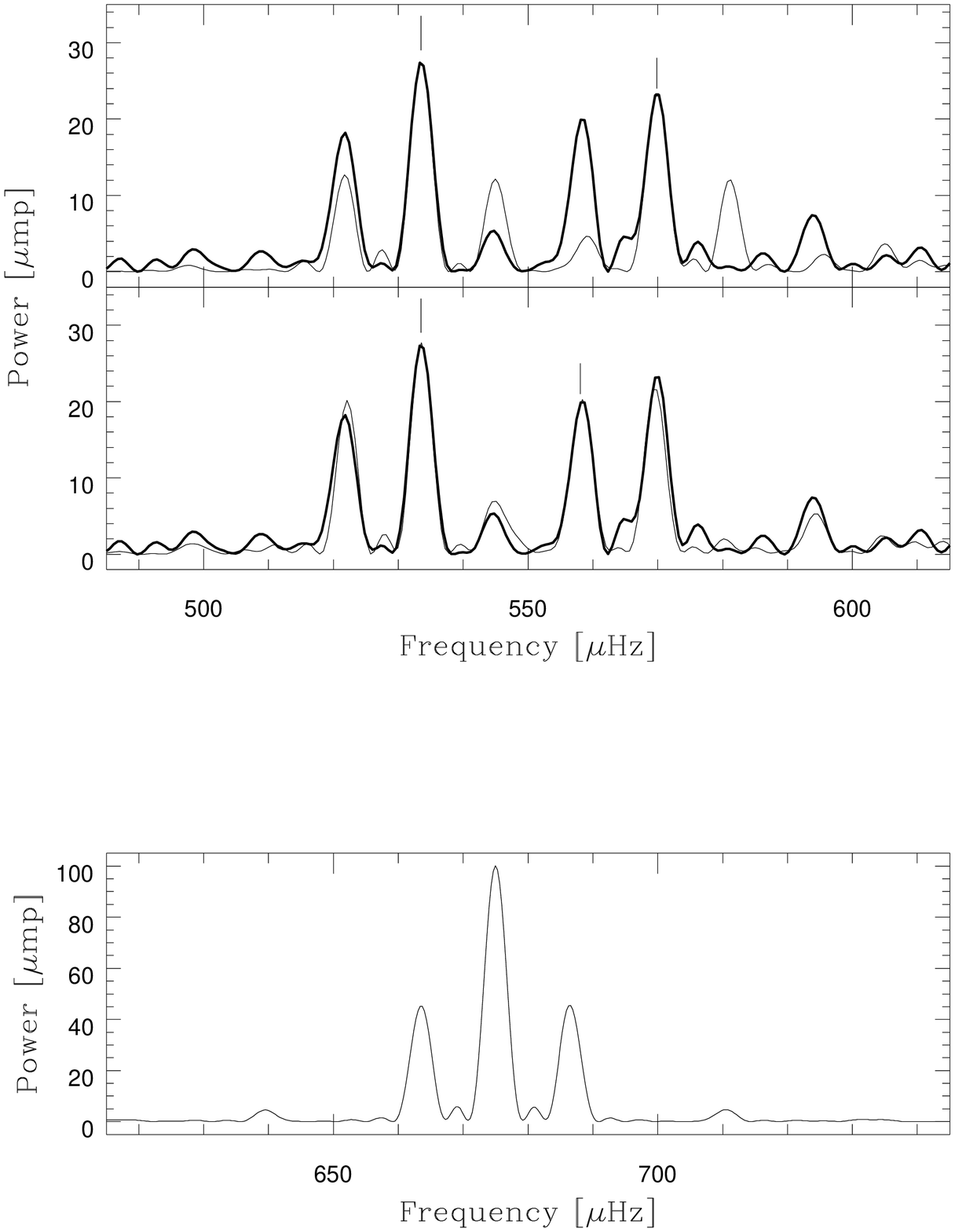]{Enlargement of the power spectrum of the 1988 data
from Kitt Peak.  The solid line in each panel is the data, and the thin lines
correspond to synthetic power spectra with the indicated input frequencies.
See text for details of the synthesis technique.  The second panel shows a
good match to the observations.  The bottom panel represents the window
function for these runs.}

\end{document}


\begin{deluxetable}{llcccc}

\tablenum{1}

\tablewidth{0pc}
\tablecolumns{6}
\tablecaption{NGC 1501 Observing Sites}
\tablehead{
\colhead{} & \colhead{} & \colhead{Pixel} & \colhead{Read Noise} &
\colhead{} &  \colhead{Exposure}  \\
\colhead{Telescope} &\colhead{Detector} & \colhead{Size}  &
\colhead{(electrons)}&\colhead{Filter}&\colhead{(s)}
}

\startdata
Kitt Peak 0.9-m &Tektronix $512\times512$ &$0\parcsec 77$ &8 &Str\"omgren $y$ &90 \nl
Lick 1-m       &GEC $288\times385$ &$0\parcsec 80$ &9 &7500 \AA &120 \nl
Okayama 0.9-m  &Hamamatsu $1024\times1024$ & $0\parcsec 20$ &15 &Johnson $V$ &120 \nl
Wise 1-m       &RCA $320\times512$ &$0\parcsec 87$ &54 &Johnson $V$ & 120 \nl
Wendelstein 0.8-m &Thomson $384\times576$ &$0\parcsec 47$ &20 &Bessell $R$ &120 \nl
\enddata
\tablecomments{
Observers were as follows: \\
Kitt Peak: Bond\\
Lick: Stover\\
Okayama: Tamura, Malasan, Yamasaki, Hashimoto, Kambe, Takeuti\\
Wise: Leibowitz\\
Wendelstein: Soffner, Roth, Mitsch\\
}
\end{deluxetable}

\begin{deluxetable}{llccc}
\tablewidth{0pt}
\tablenum{2}
\tablecaption{Table of observations of NGC 1501 for the 1991 global campaign}
\tablehead{
\colhead{Telescope} &
\colhead{Run Name} &
\colhead{Date} &
\colhead{Start Time}     &
\colhead{Run Length}     \\
\colhead{} &
\colhead{} &
\colhead{(UT)} &
\colhead{(UT)} &
\colhead{(hh:mm:ss)} 
}
\startdata
Okayama 0.9m & ok91-001 & 14 Nov 1991 & 10:11:15 & 9:12:56 \nl
Okayama 0.9m & ok91-002 & 15 Nov 1991 & 12:30:52 & 7:14:52 \nl
Okayama 0.9m & ok91-003 & 18 Nov 1991 & 13:56:26 & 5:51:11 \nl
   Wise 1.0m & wi91-001 & 18 Nov 1991 & 16:14:16 & 10:52:39 \nl
   Lick 1.0m & li91-001 & 19 Nov 1991 &  6:07:29 & 5:29:52 \nl
   KPNO 0.9m & kp91-001 & 19 Nov 1991 &  7:03:58 & 5:54:13 \nl
   Wise 1.0m & wi91-002 & 19 Nov 1991 & 16:12:40 &  8:24:15 \nl
   Lick 1.0m & li91-002 & 20 Nov 1991 &  3:04:15 & 10:24:00 \nl
   KPNO 0.9m & kp91-002 & 20 Nov 1991 &  3:25:54 & 9:15:59 \nl
Okayama 0.9m & ok91-004 & 20 Nov 1991 & 10:51:50 & 8:42:42 \nl
   Wise 1.0m & wi91-003 & 20 Nov 1991 & 16:09:46 & 10:28:57 \nl
   KPNO 0.9m & kp91-003 & 21 Nov 1991 &  3:17:32 & 9:19:01 \nl
Okayama 0.9m & ok91-005 & 21 Nov 1991 &  9:58:58 & 9:27:31 \nl
   Wise 1.0m & wi91-004 & 21 Nov 1991 & 21:12:24 &  4:09:43 \nl
   KPNO 0.9m & kp91-004 & 22 Nov 1991 &  2:42:46 & 9:51:37 \nl
   Lick 1.0m & li91-003 & 22 Nov 1991 &  8:00:35 & 5:36:00 \nl
Okayama 0.9m & ok91-006 & 22 Nov 1991 &  9:43:32 & 0:48:13 \nl
   Lick 1.0m & li91-004 & 23 Nov 1991 &  2:40:20 & 10:50:00\nl
   KPNO 0.9m & kp91-005 & 23 Nov 1991 &  3:06:07 & 9:33:27 \nl
Okayama 0.9m & ok91-007 & 23 Nov 1991 & 18:09:26 & 1:03:13 \nl
Wendelst. 0.8m & we91-001 & 23 Nov 1991 & 19:12:50 & 10:27:40 \nl
   Lick 1.0m & li91-005 & 24 Nov 1991 &  2:36:25 & 7:24:51\nl
   KPNO 0.9m & kp91-006 & 24 Nov 1991 &  2:39:44 & 10:00:14 \nl
Okayama 0.9m & ok91-008 & 24 Nov 1991 &  9:23:14 & 10:38:06 \nl
Wendelst. 0.8m & we91-002 & 24 Nov 1991 & 18:24:59 &  8:50:19 \nl
   Lick 1.0m & li91-006 & 25 Nov 1991 &  2:32:52 & 10:49:08\nl
   KPNO 0.9m & kp91-007 & 25 Nov 1991 &  3:08:32 & 9:33:40 \nl
Wendelst. 0.8m & we91-003 & 25 Nov 1991 & 16:16:09 & 12:54:58 \nl
   Lick 1.0m & li91-007 & 26 Nov 1991 &  2:28:25 & 10:50:00\nl
   KPNO 0.9m & kp91-008 & 26 Nov 1991 &  2:43:14 & 9:57:37 \nl
Wendelst. 0.8m & we91-004 & 26 Nov 1991 & 17:01:32 & 12:29:07 \nl
   KPNO 0.9m & kp91-009 & 27 Nov 1991 &  2:34:00 & 10:10:08 \nl 
\enddata
\end{deluxetable}


\begin{deluxetable}{ccccc}

\tablenum{3}

\tablecolumns{5}
\tablewidth{0pc}
\tablecaption{NGC1501 and its Comparison Stars}
\tablehead{
\colhead{Star}    &
\colhead{$\alpha$ (J2000)} &
\colhead{$\delta$ (J2000)} &
\colhead{$V$} &
\colhead{$B-V$}\\
}
\startdata
NGC 1501 &$4^{\rm h}$06$^{\rm m}$59\fs44 &+60\arcdeg$55'$14\farcs6 &
  14.20 &+0.71 \nl
Comp 1   &4 06 42.54  &+60 57 13.6   &13.05  &+0.95 \nl
Comp 2   &4 07 06.19  &+60 57 18.4   &13.81  &+0.89 \nl
Comp 3   &4 06 57.94  &+60 56 46.2   &15.45  &+1.57 \nl
\enddata
\end{deluxetable}
\begin{deluxetable}{ccccccc}

\tablenum{4}

\tablecolumns{7}
\tablewidth{0pc}
\tablecaption{Periodicities of NGC 1501 from 1991 Global Campaign Data.}
\tablehead{
\colhead{Frequency} & \colhead{Period} & \colhead{Amplitude} &
\colhead{$T_{\rm max}$\tablenotemark{a}}& comments\\
\colhead{(\uhz)} & \colhead{(s)} &\colhead{(mma)} & \colhead{(s)}
}
\startdata
  866.281 $\pm 0.048$ & 1154.36 $\pm 0.06$ & 5.00 $\pm 0.37$ & 132 $\pm 42$ 
 & \nl
  855.508 $\pm 0.130$ & 1168.90 $\pm 0.18$ & 1.84 $\pm 0.37$ & 1034 $\pm 115$ 
 & possible alias\nl
   & & & \nl
  798.777 $\pm 0.084$ & 1251.91 $\pm 0.13$ & 2.66 $\pm 0.36$ & 872 $\pm 81$ 
 & \nl
  758.463 $\pm 0.044$ & 1318.46 $\pm 0.07$ & 5.05 $\pm 0.36$ & 693 $\pm 45$ 
 & \nl
  728.366 $\pm 0.119$ & 1372.94 $\pm 0.22$ & 1.87 $\pm 0.35$ & 323 $\pm 126$
 & \nl   
 & & & \nl
  698.555 $\pm 0.124$ & 1431.53 $\pm 0.25$ & 1.80 $\pm 0.36$ & 550 $\pm 137$ 
 & \\
 & & & \nl
  661.089 $\pm 0.121$ & 1512.66 $\pm 0.28$ & 1.77 $\pm 0.36$ & 967 $\pm 146$ 
 & \\
 & & & \nl
  567.947 $\pm 0.071$ & 1760.73 $\pm 0.22$ & 3.15 $\pm 0.36$ & 102 $\pm 96$ 
 & \nl    
  528.275 $\pm 0.088$ & 1892.95 $\pm 0.32$ & 2.55 $\pm 0.36$ & 1778 $\pm 128$
 & \nl     
  500.209 $\pm 0.116$ & 1999.16 $\pm 0.46$ & 1.92 $\pm 0.36$ & 198 $\pm 178$
 & \nl
 & & & \nl   
  191.029 $\pm 0.059$ & 5234.81 $\pm 1.62$ & 3.73 $\pm 0.36$ & 4896 $\pm 238$\nl
\enddata
\tablenotetext{a}{Time of first maximum after BJDD 2448571.50367 = 0h UT 
1991 Nov 14}
\end{deluxetable}
\begin{deluxetable}{cccc}

\tablenum{5}

\tablecolumns{4}
\tablewidth{0pc}
\tablecaption{Archival Photometry of NGC 1501}

\tablehead{
\colhead{Run}&\colhead{Start Date}&\colhead{Start Time}&\colhead{Duration}\\
\colhead{name} & \colhead{(UT)} &\colhead{(UT)}&\colhead{[h:mm:ss]} \\
}
\startdata
kp87-01 & 29 Dec 1987 & 4:23:46 & 2:25:13 \nl
kp87-02 & 30 Dec 1987 & 4:23:22 & 4:49:00 \nl
kp88-01 &  2 Jan 1988 & 7:00:34 & 3:12:10 \nl
 & & & \nl
kp88-02 & 27 Nov 1988 & 3:28:17 & 9:36:59 \nl
kp88-03 & 28 Nov 1988 & 1:26:43 & 11:39:40 \nl
kp88-04 & 29 Nov 1988 & 1:21:24 & 11:32:23 \nl
 & & & \nl 
kp89-01 &  8 Oct 1989 & 5:50:55 & 6:23:22 \nl
kp89-02 &  9 Oct 1989 & 6:04:07 & 6:19:30 \nl
kp89-03 & 15 Oct 1989 & 4:59:51 & 6:40:32 \nl
kp89-04 & 16 Oct 1989 & 7:01:36 & 5:39:53 \nl
kp89-05 & 21 Oct 1989 & 8:42:10 & 2:24:34 \nl
& & & \nl 
kp90-01 &  1 Jan 1990 & 1:46:18 & 9:15:00 \nl
kp90-02 &  2 Jan 1990 & 1:17:52 & 10:01:47 \nl 
kp90-03 & 29 Nov 1990 & 2:18:41 & 11:07:24 \nl 
kp90-04 & 30 Nov 1990 & 1:42:05 & 11:36:33 \nl
kp90-05 &  3 Dec 1990 & 6:24:54 & 6:13:56 \nl
kp90-06 &  6 Dec 1990 & 7:01:34 & 5:40:38 \nl
kp90-07 &  7 Dec 1990 & 6:52:37 & 5:51:30\nl
\enddata
\end{deluxetable}
\begin{deluxetable}{ccc}
\tablenum{6}
\tablewidth{0pt}
\tablecaption{Dominant Periodicities of NGC 1501 observed in archival runs.}
\tablehead{
\colhead{Runs} &
\colhead{Frequency} &
\colhead{Amplitude} \\
\colhead{}     &  
\colhead{(\uhz)} &
\colhead{(mma)} 
}
\startdata
                   &:569.14 $\pm 0.22$ & 7.7 $\pm 1.6$ \nl
kp87-01,02,kp88-01 &:667.97 $\pm 0.21$ & 9.6 $\pm 1.6$ \nl
                   & 795.30 $\pm 0.19$ & 9.4 $\pm 1.6$ \nl
& & \nl
                 & 533.49 $\pm 0.37$ &  5.5$\pm 0.9$ \nl
                 & 558.03 $\pm 0.40$ &  5.2$\pm 0.9$ \nl
                 & 656.04 $\pm 0.18$ & 11.6$\pm 0.9$ \nl
 kp88-02-04      & 707.94 $\pm 0.34$ &  6.2$\pm 0.9$ \nl
                 & 729.46 $\pm 0.30$ &  6.9$\pm 0.9$ \nl
                 & 864.70 $\pm 0.68$ &  3.0$\pm 0.9$ \nl
                 & 900.24 $\pm 0.73$ &  2.8$\pm 0.9$ \nl
& & \nl
                 &:524.92 $\pm 1.20$ &  5.0$\pm 1.6$ \nl
                 & 557.96 $\pm 0.55$ & 10.7$\pm 1.6$ \nl
                 & 657.39 $\pm 0.61$ &  9.5$\pm 1.6$ \nl
 kp89-01,02      & 708.08 $\pm 1.40$ &  5.3$\pm 1.6$ \nl
                 & 730.40 $\pm 1.30$ &  5.9$\pm 1.6$ \nl
                 &:768.51 $\pm 1.90$ &  3.1$\pm 1.6$ \nl
                 & 833.19 $\pm 1.00$ &  5.8$\pm 1.6$ \nl
                 & 865.15 $\pm 1.00$ &  5.8$\pm 1.6$ \nl
                 & 532.46 $\pm 0.72$ &  8.5$\pm 1.8$ \nl
 kp89-03,04      & 568.12 $\pm 0.68$ &  9.0$\pm 1.8$ \nl
                 & 666.25 $\pm 0.48$ & 12.2$\pm 1.7$ \nl
& & \nl
                 & 532.50 $\pm 0.09$ & 6.2 $\pm 1.1$ \nl
                 & 558.30 $\pm 0.06$ & 8.7 $\pm 1.1$ \nl
 kp89-01-05      & 655.78 $\pm 0.05$ & 9.9 $\pm 1.1$ \nl
                 &:708.29 $\pm 0.11$ & 5.0 $\pm 1.2$ \nl
                 &:730.31 $\pm 0.11$ & 5.0 $\pm 1.2$ \nl
                 &:768.93 $\pm 0.12$ & 4.4 $\pm 1.1$ \nl
                 & 865.89 $\pm 0.09$ & 5.1 $\pm 1.1$ \nl
& & \nl
                 &:533.84 $\pm 0.78$ & 5.3 $\pm 1.1$ \nl
                 &:557.64 $\pm 0.76$ & 5.5 $\pm 1.1$ \nl
 kp90-01,02      & 655.78 $\pm 0.45$ & 9.1 $\pm 1.1$ \nl
                 & 707.85 $\pm 0.79$ & 5.2 $\pm 1.2$ \nl
                 & 728.97 $\pm 0.89$ & 4.7 $\pm 1.2$ \nl
                 & 865.87 $\pm 0.62$ & 6.5 $\pm 1.1$ \nl
& & \nl
                 & 655.75 $\pm 0.91$ & 5.1 $\pm 1.3$ \nl
 kp90-03,04      & 695.86 $\pm 0.47$ & 9.8 $\pm 1.3$ \nl
                 & 758.52 $\pm 0.61$ & 7.4 $\pm 1.5$ \nl
                 & 864.33 $\pm 0.76$ & 5.9 $\pm 1.3$ \nl
\enddata
\end{deluxetable}

\begin{deluxetable}{cccccc}
\tablenum{7}

\tablewidth{0pt}
\tablecaption{A Model for the Pulsation Spectrum of NGC 1501\tablenotemark{a}}
\tablehead{
\colhead{Frequency\tablenotemark{b}}     &  
\colhead{Period\tablenotemark{b}}     &  
\colhead{$m$}     &  
\colhead{$\Delta n$}     &  
\colhead{Model Period}     &  
\colhead{$P_{\rm model}-P_{\rm obs}$ }     \\
\colhead{(\uhzp)}     &  
\colhead{(s)}     &  
\colhead{}     &  
\colhead{}     &  
\colhead{(s)}     &  
\colhead{(s)}     
}
\startdata
500.2  & 1999 & 0  & 22 & 2003 & +4 \nl
 & & & & \nl
528.3  & 1893 &  0 & 17 & 1891 & $-2$ \nl
533.3  &      & $-1$ &    &  \nl
 & & & & \nl
558.0  &      & +1 &   \nl
(562.9)&(1776)& (0)& 12 & 1780 & +4 \nl
568.0 &       & $-1$  &    &   \nl
 & & & & \nl
656.0  &  &   &   \nl
661.1  & 1513 & 0 &  0  & 1512 & $-1$ \nl
666.3  &  &   &   \nl
 & & & & \nl
698.6 & \nl
(703.3)&(1422)& 0 & $-4$  & 1423 & +1 \nl
708.0 & \nl
 & & & & \nl
(723.4)&(1382)& (0)& $-6$ & 1379 & $-3$ \nl
728.4  &      & $-1$ & \nl
 & & & & \nl
758.5  &      & +1 & \nl
(762.5)&(1312)& (0)& $-9$  & 1312 & 0 \nl
 & & & & \nl
798.8  &      &+1 & \nl
(803.8)&(1244)&(0)& $-12$ & 1245 & 1 \nl
 & & & & \nl
866.0 & 1155 & 0 & $-16$ &  1156 & 1  \nl
\enddata
\tablenotetext{a}{$P=1512.4\,{\rm s} +22.297\,{\rm s}\times \Delta n$.}
\tablenotetext{b}{Frequencies and periods for $m=0$ modes that are inferred 
from the observed frequency spacings are indicated in parentheses.}
\end{deluxetable}